# The application of multivariate classification in evaluating the regional differentiation by population income in Russia[1]

SADOVNIKOVA Natalia A.[2]
ZOLOTAREVA Olga A.[3],

**Abstract.** The article presents the results of multivariate classification of Russian regions by the indicators characterizing the population income and their concentration. The clusterization was performed upon an author approach to selecting the characteristics which determines the academic novelty in the evaluation of regional differentiation by population income and the interconnected characteristics. The performed analysis was aimed at the evaluation of the real scale of disproportions in spatial development of the country territories by the considered characteristics. The clusterization results allowed to formulate the condition of a relatively "strong" position of a group of high-income regions (the changes in the array of regions constituting it is highly unlikely in the foreseeable future). Additionally there has been revealed a group of Russian regions that the population is struggling to live on quite low income. These so-called "poor" regions, within the crisis conditions caused by Covid-19 are in need of additional public support, without which their population will impoverish.

**Key words:** population income differentiation, population poverty, regional clusterization, Gini coefficient.

[1] Данное исследование выполнено в рамках государственного задания в сфере научной деятельности Министерства науки и высшего образования РФ на тему "Разработка методологии и программной платформы для построения цифровых двойников, интеллектуального анализа и прогнозирования сложных экономических систем", номер проекта FSSW-2020-0008.
This research was performed in the framework of the state task in the field of scientific activity of the Ministry of Science and Higher Education of the Russian Federation, project "Development of the methodology and a software platform for the construction of digital twins, intellectual analysis and forecast of complex economic systems", grant no. FSSW-2020-0008.

[2] Doctor of Economics, Professor, Head of the Department of Statistics, Plekhanov Russian University of Economics, Moscow, Russia. E-mail: Sadovnikova.NA@rea.ru
[3] Cand. Sci. (Economics), Associate professor of the Department of statistics, Plekhanov Russian University of Economics, Moscow, Russia. E-mail: zolotareva.oa@rea.ru

# Introduction

The world society has lately been focusing their sights on the problem of inequality within countries that has not evaded the attention of the United Nations Organization. Among the seventeen global aims at the area of sustainable development that must be reached until 2030, a special place is held by the goal enumerated "10": «Reducing inequality within and between countries» [1]. The realization of this goal would first be aided by the gradual achievement and support of the income growth of the least provided 40 percent population on the level above the country average.

The modern conditions of harmonization of the world and national goals the vector of Russian economy development has taken the social direction. The solution of social problems is occurring in the process of development and realization of the socio-economic policy of the state, which has, as priority, the policy of income distribution. In this context the main goal of Russian Federation is tackling the poverty and the reduction of income differentiation of the population.

Let us mark that these goals were defined more than a decade ago but they have not yet lost their relevance in the present day, as their resolution in the conditions of internal and external shocks appears a hard enough task. Presently there is carried out the review of many strategic documents that are connected, with, e.g., the closure of the Concept of long-term socio-economic development over the period of until 2020, approved by the order of the Government of the Russian Federation of November 17, 2008 N 1662-p.[2] One of the priorities of regional development established by the Concept-2020, was declared as securing the balanced socio-economic development of the country regions.

In his message to the Federal Assembly in 2018, the President of Russia has clearly established certain priorities, calling "a person, his present and future, is the main meaning and goal of our development." These words by V.V. Putin were reinforced by the May decree, which posed a difficult task for the Government - to preserve and increase human capital through social, economic and infrastructure policies.

Among the goals specified in this document there have been highlighted the following: ensuring sustainable growth in real incomes of citizens, as well as halving the poverty level in the Russian Federation [3]. According to the head of our state, the implementation of these measures could allow for a breakthrough socio-economic development of Russia, improve the living standards of citizens and create the conditions for their self-realization and the realization of the talent of each person.

Presently, the national goals specified in the May decree have been corrected, and already in the Decree of the President of the Russian Federation of July 21, 2020 No. 474 "On the national development goals of the Russian Federation for the period up to 2030" as a target parameter characterizing the achievement of such a strategic goal as "Preservation of the population, health and well-being of people", an indicator is given: "A reduction in the poverty level by half compared to the 2017 level." [4] Moreover, an array of other indicators have undergone correction in the context of: to enter not the top five but the top ten; to achieve a goal not by 2024, but by 2030, et cetera. It is inarguably important that the analysis and the update of goals have been performed. Presently we see the main improvement in the area of administration, which is its purposefulness / goal-setting. It is only a certain stage of the work, according to which an initiative is needed for changes / updates / additional development (if necessary) of national projects, and it must be carried out with the account of expert evaluations and the suggestions of academic society. Let us note, according to Dmitry Sorokin the Corresponding Member of the Russian Academy of Sciences: "The technological breakthrough and poverty are mutually exclusive things, this is the trivia of economic science," the scientist emphasized, "therefore, the first national project should be the fight against poverty." [5] However, there is currently no national project to combat poverty at all.

The above-stated defines high practical significance of statistical evaluation of regional differentiation of Russia at the current stage of country development, the data upon which must be set as a basis of informational and analytical support for making coordinated decisions on the federal, regional and local levels on reducing the degree of interregional differentiation and the development of the economy of each region.

**Methodology**

The global challenges of the XXI century within the conditions of humanitarian-technological revolution have defined the vector of changing the paradigm of socio-economic development from "person for the economy" towards "economy for the person". In this context the quality of life of the population appears to be one of the priority indicators positioning the state in the global environment. Among the life quality indicators there are specified the indicators of income, level of poverty and the Gini coefficient. The monetary income defines the satisfaction degree of personal needs, and allows to evaluate the material welfare of the population. The monetary income includes the wages for labor for all population categories, pensions and allowances, scholarships and other social transfers, the receivables of agricultural products sales, the property incomes in the form of interest on deposits and securities, the dividends, the incomes of individual entrepreneurs, as well as insurance claims, income from the sale of foreign exchange and other income [6].

Presently the generation of income is carried out within the conditions of deep regional disparity of Russia, that is caused by the existence of a conglomerate of factors, among which the following must be listed: [7]

- the huge space of the territory with extremely uneven placement and highly differentiated physical density of the citizens, for which about 80% population is concentrated within the Central European part of the country while the major part of minerals and energy resources are located beyond Ural;
- the significant and potentially dangerous differentiation in the Russian Federation regions by the level of socio-economic development, that periodically causes centrifugal currents, and demands the accounting of local specifics in making every administrative decision.

- the complicated demographic situation that spawns severe problems of labor force reproduction, migration expansion and gradual extinction of separate regions of the country including the strategically important ones.

This demands the evaluation of regional differentiation by the population income as well as by poverty level and the Gini coefficient that characterizes the degree of social delamination of the society as it represents the relation among average levels of monetary income for the 10% population with the highest incomes to the 10% population with the lowest incomes.

Thus the substantiation has been established for singling out the three key characteristics of regional differentiation in population income in Russia. The instrument that allows to effectively resolve this task is the cluster analysis, as the clusterization in particular enables to classify the studied objects characterized by a number of factors, into uniform apriori-undetermined target groups and analyze a significantly large amount of information and represent the data arrays in a more demonstrative way. This confirms the result of content-analysis of the clustering procetures in the modern statistical studies, carried out in the work of Ilyshev M.M. and Shubat OM, which made it possible to assert: "... the practice of applying multidimensional methods to various socio-economic aggregates has shown their effectiveness and the possibility of obtaining highly analytical scientific and practical results obtained in the course of economic and statistical research." [8]

**Results**

The clusterization of regions of Russia by level of financial welfare of the population of Russia was performed by the 2018 time slice upon the three factors:

$x_1$ – average population monetary income per capita, rub.,

$x_2$ – proportion of population with monetary income below living wage, %;

$x_3$ – Gini coefficient, factor/times.

The data for clustering were taken for the year 2018 which is because there are presently absent the disseminated data upon the three specified factors for 2019.

The R-STUDIO software package was used for the processing and analysis of the statistical information.

The analysis of squared distances within the clusters has enabled to state the need of dividing the regions of Russia into 4 homogeneous groups by the level of financial well-being of the population.

The clustering of Russian regions upon the nearest-neighbor method has also allowed to conclude that the studied administrative territorial units must be organized into four clusters (Fig.1).

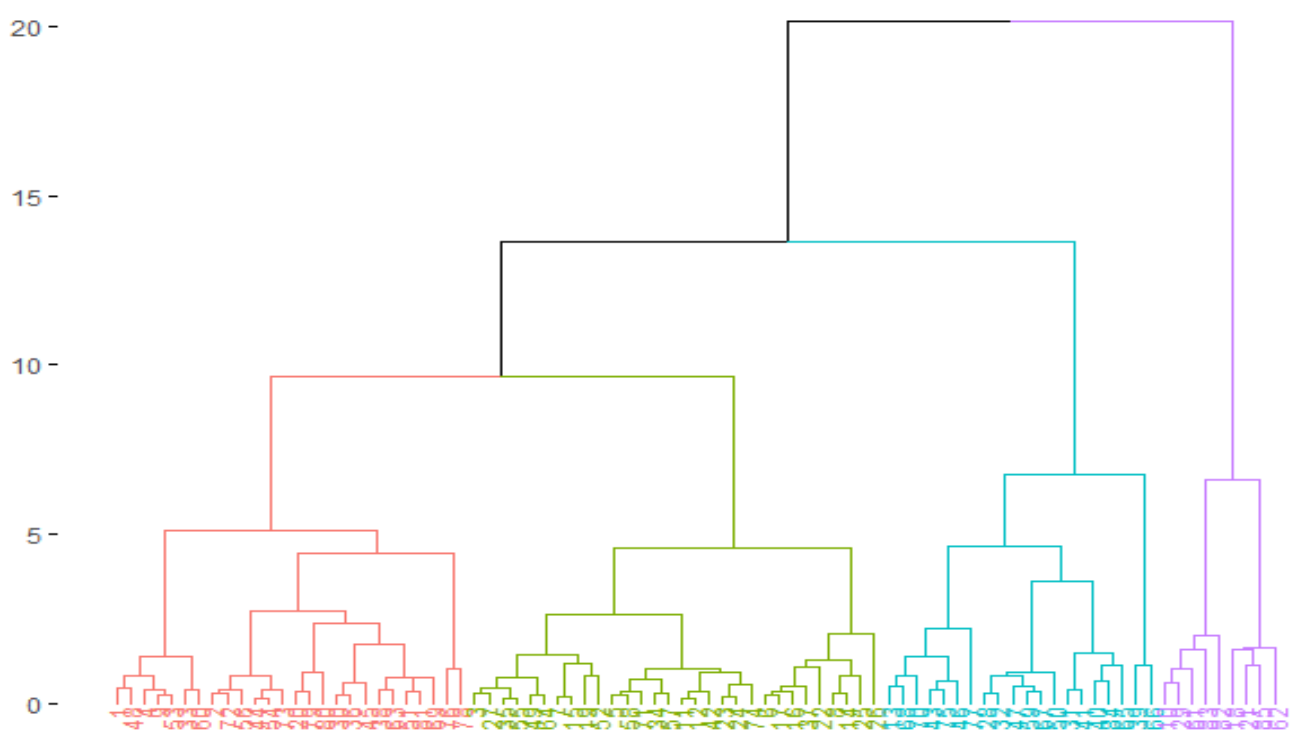

Figure 1 – The clusterization of Russian regions by the differentiation of monetary income in 2018, nearest neighbor method

The clusterization quality upon the nearest neighbor method is confirmed by the obtained correlation coefficient value of 0,951.

Upon the k-means method, eighty-five regions of the Russian Federation were divided into four clusters differentiating by the indicators that characterize the financial well-being of the population in 2019. The first group includes 10 regions of the Russian Federation, the second group includes 8, the third group includes 31 and the fourth group includes 35 administrative regional units.

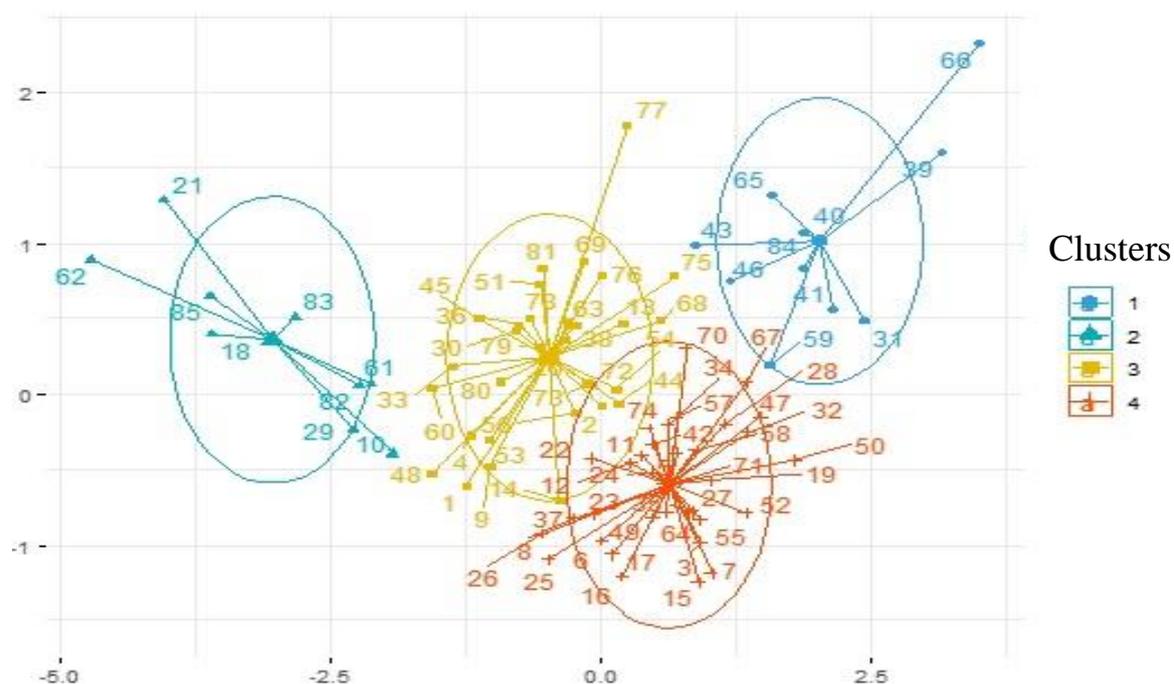

Figure 2 – The clusterization of Russian regions by the level of monetary income differentiation in 2018, k-means method.

The quality of the method is confirmed by the explained variance percentage equalling 72.5%. It can be stated that the regions of the country have been significantly differentiated by the level of population income, which is evidenced by high dispersion of means by the cluster indicators (Table 1).

Table 1 – The average levels of indicators describing the clusters unifying the regions of Russia by the degree of monetary income differentiation in 2018

| Cluster | Number of regions | Average monetary income per capita, rub. | Proportion of the population with monetary income below living wage, % | Gini coefficient, times |
|---|---|---|---|---|
| First | 10 | 19 521 | 24,37 | 0,355 |
| Second | 9 | 62 125 | 8,00 | 0,409 |
| Third | 31 | 30 196 | 13,50 | 0,389 |
| Fourth | 35 | 25 308 | 13,70 | 0,355 |

The first group is composed of the regions of Russia with impoverished population: Republic of Kalmykia; The Republic of Ingushetia; Kabardino-Balkar

Republic; Karachay-Cherkess Republic; Chechen Republic; Mari El Republic; Kurgan region; Altai Republic; Tuva Republic and Jewish Autonomous Region. The population of the 1st cluster regions in 2018 is characterized by the lowest level of average personal population income. The average level of the studied indicator was 19 521 rub., which is 3.2 times less the analogical characteristic for the 2nd cluster regions, where the average personal income amounted to 62 125 rub. (Table 1).

Simultaneously, the population of the poorest regions of Russia is characterized by the highest proportion of the persons with monetary income below living wage that was 24.37%. This indicator is lower than the analogical for the regions with the most provided population by 3.0 times.

The differentiation of income of the least provided citizens is relatively low, the Gini coefficient was 0.355 times which is equivalent to the analogical indicator of the fourth cluster regions with average financial well-being. The Gini coefficient for the most provided population group was 0.409 times, which is 15.2% greater than for the 'poorest' regions of the Russian Federation.

Let us note that the available data for 2019 allow to state that these regions, for their most part, will be characterized as "socially unprovided". In 2019 the average per-capita monthly income per person in these regions fluctuates between 16497 rub. (Tyva Republic) to 26099 (Jewish Autonomous Region), while the average income for Russia in overall is 35 188 rubles. The poverty of such regions as Karachay-Cherkess Republic, Republic of Kalmykia, Republic of Altai, Republic of Tyva is explained by the insufficiency of support of their economic development: the industrial production index in 2019 was, in percent to 2018, correspondingly for these regions: 100,6 %; 97,3 %; 89,3 %; 99,7 (with 102.4% for Russia in overall); the proportion of investments in capital assets was 0.1% of total investments in Russia in overall.

The second cluster has unified the least number of the regions of Russia, that can be characterized as the most financially provided: Moscow and the Moscow region (constituting the Moscow agglomeration), St. Petersburg, Khanty-Mansi, Yamalo-Nenets, Nenets, Chukotka autonomous districts, Magadan and Sakhalin regions. In these regions the average population income in 2018 was 62 125 rub. The value of this indicator is 2.1 and 2.5 times greater than the analogical indicators for the regions with the level of financial well-being of above average and of average level (30 196 rub. and 25 308 rub., corresponsingly). This corresponds to the economy of the regions (the Moscow agglomeration and the city of St. Petersburg as two capital regions, and gold, oil, gas and coal-extracting regions), and the high level of population employment.

The average proportion of people below the living wage was minimal in the most provided regions of Russia: 8.00%, which is 1.7 times lower than the analogical indicators of the third and the fourth clusters.

The differentiation of the population by financial well-being in the wealthiest regions of Russia was the highest and was equal to 0.409 times. The Gini coefficient for the second cluster regions is 5.15% greater than the analogical indicator for the regions which population is characterized by above-average financial well-being (0.389 times).

The third group of Russian Federation subjects is constituted by the Lipetsk region, the Smolensk region, Krasnodar Territory, the Republic of Bashkortostan and some other territories. The average income of the regions with the above-average financial well-being was 30 196 rub. in 2018, which is greater than the analogical indicator for the fourth group by 4 888 rub. (by 19.32%).

The proportion of Russian regions of the third cluster with the income below the living wage was 13.50%. This indicator is lower than the level of poverty for the least financially provided regions of the Russian Federation by 0.2 pp (by 1.53%).

The population of Russian regions with the above-average financial well-being is characterized by a relatively high differentiation of income. The Gini coefficient is the second highest for these regions, its value being 0.389 times which is 9.61% greater than the analogical indicator for the fourth group regions.

As part of the largest fourth cluster, there are the Oryol, Tver and the Volgograd regions, the city of Sevastopol and many other regions of Russia. The value of average per-capita population income in these regions was 25 308 rub. in 2018, which is above the per-capita income for the least provided regions of the country by 5 787 rub. (by 29.60%).

About 13.70% citizens living in the regions of the analyzed cluster received, in 2018, the personal income below the living wage. This indicator takes the second place among all obtained clusters and is lower than the analogical indicator for the population of the 'poorest' regions of the country by 77.88%.

The amount of income for the population residing on the Russian regions territory with the average level of financial welfare is distributed evenly. For the regions of the fourth cluster the Gini coefficient takes a low value (0.355 times). The studied indicator is equal to the Gini coefficient obtained for the regions of the first cluster.

## Conclusions

While generalizing the study results, it is possible to state the following conclusions.

The Russian Federation regions are different in terms of population income, and it appears expedient to divide them into four clusters, including the group of the most financially provided regions and the "poor" regions group, or "socially disadvantaged" regions.

The cluster unifying the regions with the wealthiest population of the country includes the cultural centers (the cities of Moscow and St. Petersburg), as well as the largest centers producing the export-meant output (Yamalo-Nenets and Chukotka Autonomous Districts, Magadan Region). For these regions the average population income is above its level for the whole country in total, and the concentration of population with income below the living wage is lower than the analogical characteristic not only for other clusters but also lower than for Russia in overall, which can be explained by the effect of regional coefficients for salary of workers. However, the Gini coefficient in the regions of these cluster is the highest compared to other regions. The specified cluster is relatively "solidified" by the level of income, and the changes in its composition is hardly possible in foreseeable future.

The "poor" or "socially disadvantaged" regions cluster includes the regions of Russia for which the average personal income is lower than the country level in 2018. These Russian Federation regions have the maximum number of people with incomes lower than living wage, but the subjects of these regions are characterixed by the lowest differentiation of income which is explained by the low earnings of the residents of these regions.

The third and the fourth clusters containing over ¾ regions of the country can be named the regions of Russia with average financial well-being. The values of such characteristics as average per-capita income and the proportion of population below the living wage in these regions are the closest to the analogical characteristics for the Russian Federation in overall. Both identified clusters attest that within the current conditions connected with the humanitarian technological revolution it is hard to assume that one or another region will be either the leader or an outsider among these groups. The regions in these clusters may move from one to another depending on how effectively they will use their socio-economic potential in the future.

The study indicated that the differentiation of population income within the regions with low incomes is insubstantial, while it is maximal within the regions with the highest earnings. With the growth of average per-capita income within the population of one or another region the growth of their differentiation will also be traced, that is, the wealthy will become wealthier while the poor will become even more impoverished. Moreover, the wealthy regions will continue to accumulate wealth (or remain the wealthiest among other), and the poor regions will become poorer. This tendency escalates in the conditions of the current crisis caused by the pandemic, which defines the need to develop additional measures to support the "poor" or "socially disadvantaged" regions, whose budgets are unable to withstand the present challenges.

The amount of average per-capita income in 2018 was 33 178 rub. in overall in Russia. This incidator is 28 497 rub. lower (by 87.25%) than the average population income for the ten regions of the wealthiest cluster, and higher than the analogical indicators for the regions in other clusters.

Consequently, achieving the proportional distribution of income on the country territory appears impossible due to the effect of institutional economy and inherently different resourse capacity of each region.

**Bibliographic references**

2020. DOI: 10.35808 / ersj / 1664. - [Electronic resource]. - Access mode: https://www.ersj.eu/journal/1664 (date of access: 09/10/2020).